\documentclass[aps,prl,final,twocolumn,letterpaper]{revtex4-1}

\usepackage{graphicx}   
\usepackage{import}                         
\usepackage{epstopdf}
\usepackage{amsmath} 
\usepackage{float} 
\usepackage{bm}
\usepackage{amssymb}
\usepackage{quotes}
\usepackage{indentfirst}
\usepackage{color}
\usepackage{transparent}
\usepackage{dcolumn}
\usepackage{braket}
\usepackage{multirow}
\usepackage{cancel} 
\usepackage{mdframed}
\usepackage{color}
\usepackage{bm}
\usepackage{dsfont}
\usepackage{slashed}
\usepackage{soul, color} 
\soulregister\ref{7}  
\soulregister\cite{7} 
\renewcommand{\st}[1]{}

\usepackage{xr}
\newcommand{\Lu}{$\textrm{Lu}_2\textrm{O}_3$(Eu)}
\makeatletter
\newcommand*{\addFileDependency}[1]{
  \typeout{(#1)}
  \@addtofilelist{#1}
  \IfFileExists{#1}{}{\typeout{No file #1.}}
}
\makeatother

\newcommand*{\myexternaldocument}[1]{%
    \externaldocument{#1}%
    \addFileDependency{#1.tex}%
    \addFileDependency{#1.aux}%
}

\myexternaldocument{si}

\usepackage{textcomp} 
\usepackage{xifthen}
\usepackage{xcolor}
\usepackage{etoolbox}
\newboolean{togglechanges} 
\usepackage{siunitx}
\usepackage{wrapfig}

\setboolean{togglechanges}{false}

\newcommand{\comment}[1]{\ifbool{togglechanges}
    {#1}  
    {\textcolor{blue}{#1}}}

\usepackage{bibentry}

\usepackage{graphicx}
\usepackage{dcolumn}
\usepackage{bm}
\setcitestyle{numbers,square,super}

\usepackage{textcomp} 
\begin{document}
\rmfamily

\title{Supercollimating photonic crystal scintillators}
\author{Seou~Choi$^{\bigstar,1}$}
\email{seouc130@mit.edu}
\author{Sachin~Vaidya$^{\bigstar,1,2}$}
\email{svaidya1@mit.edu}
\author{Charles~Roques-Carmes$^{1,3}$}
\author{Marin~Solja\v{c}i\'{c}$^{1,2}$}
\affiliation{$^\bigstar$ denotes equal contribution.\looseness=-1}
\affiliation{$^{1}$ Research Laboratory of Electronics, Massachusetts Institute of Technology, Cambridge, MA 02139, USA\looseness=-1}
\affiliation{$^{2}$ Department of Physics, Massachusetts Institute of Technology, Cambridge, MA 02139, USA\looseness=-1}
\affiliation{$^{3}$ E. L. Ginzton Laboratories, Stanford University, Stanford, CA 94305, USA\looseness=-1}

\clearpage 

\setlength{\parskip}{0em}
\vspace*{-2em}


\vspace{0.8cm}

\begin{abstract}

   Scintillators convert X-ray energy into visible or near-visible photons, enabling applications in high-energy particle detection and X-ray imaging. Increasing scintillator thickness improves X-ray absorption but degrades spatial resolution due to diffraction-induced lateral spreading of emitted light, resulting in a fundamental trade-off between detection efficiency and image resolution. Here, we propose a class of three-dimensional photonic crystal scintillators that overcomes this limitation through supercollimation, in which light propagates with suppressed diffraction. We develop a multiscale modeling framework that integrates nanophotonic band-structure simulations with Monte Carlo particle transport to quantitatively evaluate the performance of such scintillators. Our results show that supercollimating photonic crystal scintillators can enhance spatial resolution by up to an order of magnitude relative to conventional bulk scintillators of equal thickness. This improvement leads to a substantial increase in detector quantum efficiency (DQE), particularly at high spatial frequencies, enabling fine features to be preserved at reduced X-ray dose. We further demonstrate that comparable image quality can be achieved with approximately an order-of-magnitude lower X-ray dose. By directly engineering light transport within the bulk of the scintillator, this work establishes a nanophotonic route to simultaneously improving resolution and dose efficiency in X-ray imaging.

\end{abstract}

\maketitle

\section*{Introduction}
Scintillation is the process by which a material emits visible or near-visible light via spontaneous emission following X-ray absorption. By converting X-rays into detectable optical signals, scintillators find uses in a broad range of applications, including high-energy particle detection~\cite{moses2002current}, homeland security~\cite{gektin2017inorganic}, and medical imaging~\cite{van2002inorganic}. Efficient X-ray absorption generally requires scintillators with thicknesses comparable to or exceeding the material attenuation length. However, as thickness increases, the emitted light undergoes diffraction-induced lateral spreading, leading to degradation of spatial resolution. This establishes a fundamental trade-off between detection efficiency (brightness) and spatial resolution in scintillator-based imaging systems. In medical imaging, for example, this trade-off directly impacts patient radiation dose: maintaining high spatial resolution often necessitates increased X-ray exposure to compensate for optical blurring and noise. Given the widespread use of radiography for medical imaging and its association with non-negligible projected lifetime cancer risks from ionizing radiation~\cite{smith2025projected}, reducing dose without compromising image quality remains an important challenge.

Nanophotonic engineering has emerged as a promising strategy for tailoring the emission and propagation properties of the scintillation light~\cite{singh2024bright,carr2024toward}.  Early approaches primarily focused on enhancing light extraction efficiency by patterning nanostructures at the surfaces of bulk scintillators. Two-dimensional photonic crystal slabs~\cite{roques2022framework,martin2025large} and microlens arrays~\cite{yuan2020directional} have been shown to improve the outcoupling efficiency of scintillation light by extracting light that would otherwise be trapped by total internal reflection. Despite their effectiveness, these approaches are largely restricted to low-dimensional, surface level modifications and do not alter the intrinsic spontaneous emission process of the bulk scintillator. Recent efforts have explored fully nanostructured scintillators in which the electromagnetic environment is engineered at length scales comparable to the emission wavelength to directly modify spontaneous emission through the Purcell effect~\cite{purcell1946spontaneous}.  Photonic crystal cavities~\cite{ye2022enhancing}, one-dimensional photonic crystal slabs~\cite{kurman2024purcell, kurman2020photonic}, two-dimensional photonic crystal slabs~\cite{roques2022framework}, bulk-patterned photonic crystals~\cite{jurgensen2025volumetrically}, and nanoplasmonic structures~\cite{ye2024nanoplasmonic} have demonstrated enhanced scintillation yield and accelerated scintillation decay through tailored photonic density of states. Nevertheless, even in these platforms, the design objective has predominantly been emission rate or brightness enhancement rather than controlling the directional spatial transport of scintillation light through the photonic dispersion of the structure. Such a capability could directly address the resolution-efficiency trade-off, leading to significant improvements for imaging applications~\cite{shultzman2023enhanced}.

Here, we introduce a new class of nanophotonic scintillators--supercollimating photonic crystal scintillators--that are capable of mitigating this universal resolution-efficiency trade-off. We develop a multi length- and energy-scale modeling framework to explain how bulk energy deposition from high-energy ionizing radiation governs localized dipole emission of scintillating light in three-dimensional photonic crystals. We achieve this by combining a first-principles description of dipole emission in photonic crystals with Monte Carlo simulations of X-ray energy deposition and image formation. Using this framework, we show that by engineering photonic bands with flat isofrequency contours, scintillation light can propagate with suppressed diffraction, leading to substantially reduced lateral spreading compared to bulk scintillators. This mechanism enables up to an order-of-magnitude improvement in spatial resolution for scintillators of equal thickness. By comparing two realistic material platforms with different supercollimation bandwidths, we find that narrower emission spectra yield stronger enhancements in the modulation transfer function (MTF), with corresponding gains in detector quantum efficiency (DQE), particularly at high spatial frequencies. As a result, fine image features are preserved with significantly reduced X-ray dose. To illustrate the implications for biomedical imaging, we simulate X-ray imaging of a human kidney vasculature and show that supercollimating photonic crystal scintillators can achieve comparable image quality at approximately an order-of-magnitude lower dose while resolving fine features on the order of \SI{100}{\micro\meter}. These results establish dispersion engineering of scintillators as an important route towards high-resolution, low-dose X-ray imaging.

\section*{Modeling supercollimation in photonic crystal scintillators}

The propagation of scintillation light is fundamentally governed by the dispersion relation of the medium. In regions where the dispersion surface is locally flat, light can propagate with suppressed diffraction while maintaining its directionality over extended distances~\cite{kosaka1999self, rakich2006achieving, lu2006experimental, joannopoulos2008molding}. This phenomenon, known as supercollimation, has been extensively studied in photonic crystals for applications such as optical routing~\cite{prather2004dispersion}, sensing~\cite{wang2012photonic}, and enhanced free-electron radiation~\cite{yang2023photonic}.  Translating this concept to scintillators offers a pathway to suppress the optical blurring inherent to bulk materials, enabling simultaneous improvement in spatial resolution and detection efficiency. However, prior modeling methods are insufficient to study supercollimation in photonic crystal scintillators, as they either rely on selective excitation of flat-band photonic states using external sources~\cite{prather2004dispersion, wang2012photonic,yang2023photonic} or fail to capture the distributed, stochastic nature of scintillation under X-ray excitation~\cite{yasar2021spatially}. 

\begin{figure}[t]
    \centering
    \includegraphics[scale = 1.1]{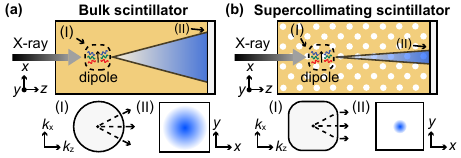}
    \vspace*{-4mm}
    \caption{\textbf{High spatial resolution X-ray imaging with supercollimating photonic crystal scintillators.} (\textbf{a} and \textbf{b}) Scintillation in (\textbf{a}) a bulk scintillator and (\textbf{b}) a supercollimating scintillator. (\textbf{a}) High energy particles such as X-rays create a dipole that emits visible light. Scintillating light propagates with lateral spreading, due to a circular isofrequency contour (IFC) (panel I), reducing the spatial resolution of X-ray images (panel II). (\textbf{b}) Tailoring the IFC to have a flat surface toward the back plane of the scintillator, scintillating light can be collimated (panel I), enhancing the spatial resolution of X-ray images (panel II). 
}
    \label{fig:concept}
\end{figure}

Here we introduce a complete description that accounts for the full set of processes from X-ray energy deposition to photon transport and detection. Fig.~\ref{fig:concept} illustrates the concept of a supercollimating photonic crystal scintillator. Upon X-ray absorption, an energy cascade ensues, eventually leading to excitations that relax via radiative pathways (e.g., creating electron-hole pairs that recombine and emit visible light in crystalline scintillators)~\cite{klein1968bandgap}. The angular emission of a dipole emitter at a particular frequency is determined by the isofrequency contour (IFC) which consists of a set of wave vectors $\textbf{k}$ with a frequency $\omega$, derived from the dispersion relation of the surrounding medium $\omega(\textbf{k})$. The emitted waves propagate normal to the IFC at each wave vector $\textbf{k}$, with a localized dipole source generating waves with all $\textbf{k}$, weighted by the local density of states~\cite{joannopoulos2008molding}. 

In a homogeneous medium such as a bulk scintillator, the IFC forms a circle (Fig.~\ref{fig:concept}(a), panel I). Therefore, the emitted light with different $\textbf{k}$ components propagates in all directions, which is simply the phenomenon of classical diffraction. This causes the lateral spreading of the propagating beam (Fig.~\ref{fig:concept}(a), panel II), resulting in decreased spatial resolution at a detector as the thickness of the scintillating material is increased. The situation changes when we engineer the dispersion relation of the photonic crystal scintillator to have a flat IFC (Fig.~\ref{fig:concept}(b), panel I). In this case, light emitted from the dipole with different $\textbf{k}$ components can be strongly collimated, preserving the spatial resolution of the X-ray image at large scintillator thicknesses (Fig.~\ref{fig:concept}(b), panel II).

The angular emission characteristics of the supercollimating photonic crystal scintillators can be calculated by assuming that the incoming X-rays generate a laterally uniform density of dipole emitters with random orientations. The emission rate of a dipole emitter, $\Gamma(\textbf{k};\omega,n)$, into a band $n$ at a certain wave vector $\textbf{k}$ and frequency $\omega$ is defined using the group velocity $v_g(\textbf{k};\omega,n)$ and the area of the IFC $A (\textbf{k};\omega,n)$:

\begin{equation}
\begin{aligned} \label{eq:coupling rate}
    \frac{d\Gamma(\textbf{k};\omega,n)}{dA (\textbf{k};\omega,n)} = \frac{C}{ v_g(\textbf{k};\omega,n)}
\end{aligned}
\end{equation}
In this expression, $C$ is the constant determined by material properties of the scintillator. Details on the derivation of Equation~\ref{eq:coupling rate} can be found in the Supplementary Note 1. 

\begin{figure*}
    \centering
    \includegraphics[scale = 1.0]{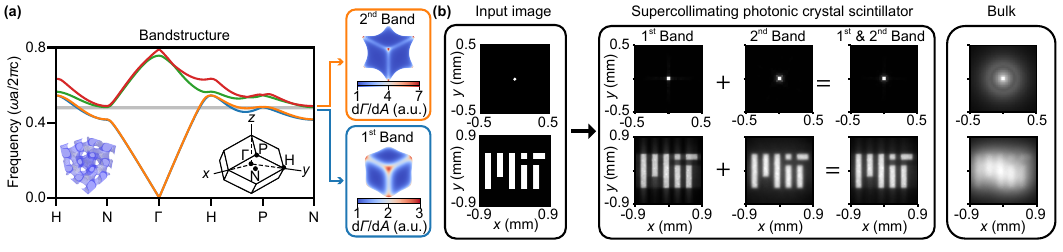}
    \vspace*{-4mm}
    \caption{\textbf{Supercollimation in photonic crystal scintillators.}  (\textbf{a}) Band structure of the three-dimensional single gyroid photonic crystal scintillator calculated using MIT Photonic Bands (MPB)~\cite{johnson2001block}. Insets show the photonic crystal structure and the first Brillouin zone. Isofrequency contours (IFCs) of the first two bands are shown in the right panel. (\textbf{b}) X-ray sources shaped through a small pinhole and the MIT logo are used (left panel). X-ray images generated by the photonic crystal scintillator (middle panel) and the bulk scintillator (right panel) of equal thickness are shown. The thickness of the scintillator is \SI{300}{\micro\meter}. Refractive index, scintillation yield and X-ray attenuation are modeled with \Lu~scintillators.}
    \label{fig:single_frequency}
\end{figure*}

With this in mind, we consider concrete photonic crystal scintillator structures that exhibit supercollimation. In particular, we examine a three-dimensional single gyroid photonic crystal made out of \Lu ~scintillator (refractive index: 1.95). This structure exhibits fully closed, flat IFCs for its bottom two bands in the frequency range highlighted with a gray area in the band structure shown in Fig.~\ref{fig:single_frequency}(a). The right panel of Fig.~\ref{fig:single_frequency}(a) shows the IFCs with each $\textbf{k}$ colored according to the value of $d\Gamma/dA$, which increases near the vertices and the edges of the IFC as the group velocity decreases near these points. Although the light emission from these areas causes unwanted blurring, the contribution from these points is insignificant as the area of these points is small compared to the total area of the IFC. From Fig.~\ref{fig:single_frequency}(a), we extract the angular distribution of the scintillating light propagating inside the supercollimating photonic crystal scintillator. Details on the simulation for sampling the angular emission can be found in the Methods section.

\section*{Spatial resolution enhancement}
We simulate X-ray imaging using such a photonic crystal scintillator with Geant4, which implements a Monte Carlo method for simulating the transport of high-energy particles~\cite{agostinelli2003geant4}. The left panel of Fig.~\ref{fig:single_frequency}(b) shows two types of input X-ray sources that we image (pencil beam and MIT logo) using the supercollimating photonic crystal scintillator with a uniform scintillator made out of the same material serving as a reference. When a scintillation event occurs within the supercollimation frequency bandwidth (i.e., gray shaded area in Fig.~\ref{fig:single_frequency}(a)), its direction of emission is stochastically sampled from the IFCs of the two bands (calculated from nanophotonic simulations), with its probability determined by the sampling weight as given in Equation~\ref{eq:coupling rate}. The middle panel of Fig.~\ref{fig:single_frequency}(b) shows that X-ray images generated from such structures exhibit very high spatial resolution as compared to the reference case with the same thickness. Details on the X-ray imaging simulation using Geant4 can be found in both the Methods section and the Supplementary Note 2. 

\begin{figure*}
    \centering
    \includegraphics[scale = 1.0]{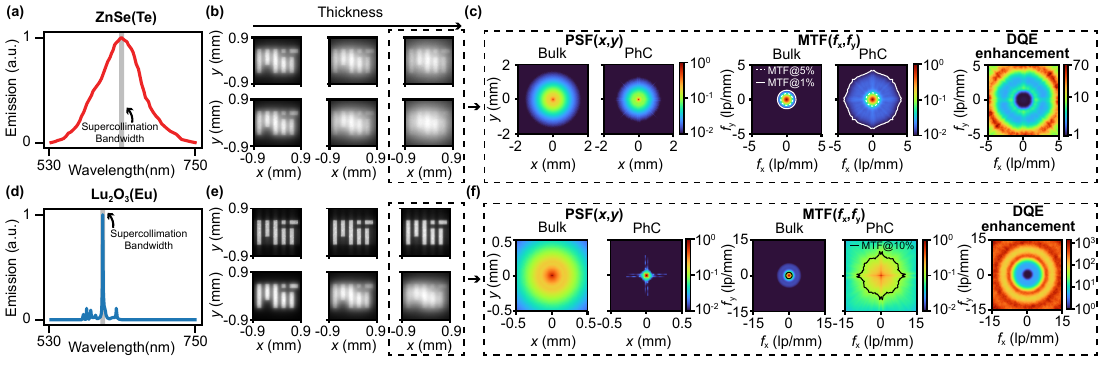}
    \vspace*{-4mm}
    \caption{\textbf{Detector Quantum Efficiency (DQE) enhancement in supercollimating photonic crystal scintillators.} (\textbf{a} and \textbf{d}) Emission spectrum and the supercollimation bandwidth for (\textbf{a}) ZnSe(Te) and (\textbf{d}) \Lu. (\textbf{b} and \textbf{e}) Simulated X-ray images for different thicknesses of the scintillator. The upper (lower) row of each sub-figure shows X-ray images of the MIT logo generated by supercollimating photonic crystal scintillators (uniform scintillators). The thickness of the scintillator in the three panels is $1.5\mu, 3.0\mu$ and $4.5\mu$, where $\mu$ is the attenuation length of each scintillator ($\mu \sim$  \SI{130}{\micro\meter} for ZnSe(Te) and $\mu \sim$ \SI{60}{\micro\meter} for \Lu~at X-ray energy of \SI{30}{\kilo\eV}).  (\textbf{c} and \textbf{f}) Point spread function (PSF), modulation transfer function (MTF) and detector quantum efficiency (DQE) enhancement of each scintillator. DQE enhancement is defined as the (DQE of the supercollimating photonic crystal scintillator)/(DQE of the bulk scintillator).} 
    \label{fig:multiple_frequency}
\end{figure*}

We next calculate and quantify the spatial resolution enhancement from these structures. For this, we consider two realistic scintillating materials with similar peak emission wavelengths but different emission bandwidths (Fig.~\ref{fig:multiple_frequency}(a) and (d)):  ZnSe(Te) and \Lu~(material properties from Ref.~\citenum{linardatos2020optical},~\citenum{sengupta2015bright}, respectively). The structure of each photonic crystal scintillator is optimized to maximize the frequency bin having two flat IFCs as in Fig.~\ref{fig:single_frequency}(a). While more than 50\% of the \Lu~emission spectrum stays inside the supercollimating bandwidth (gray shaded area), it is only $\sim$9\% for ZnSe(Te). Although some collimation will be retained immediately outside the supercollimation bandwidth, here we assume that light outside this bandwidth is emitted isotropically, capturing a worst case scenario. Under these considerations, Fig.~\ref{fig:multiple_frequency}(b) and (e) show improvements in the spatial resolution of the X-ray images over a uniform scintillator. We find that for both ZnSe(Te) and \Lu, the improvement becomes significant as the thickness increases, particularly for the \Lu ~scintillator. 

To evaluate the spatial resolution improvement, we calculate the modulation transfer function (MTF) of the photonic crystal scintillator and the bulk scintillator. MTF quantifies the extent to which the contrast of an input image is preserved along different spatial frequencies as it passes the imaging system, which can be calculated from the Fourier transform of the point spread function (PSF) (Fig.~\ref{fig:multiple_frequency}(c) and (f)). Spatial resolution is defined as the highest spatial frequency at which the MTF falls to a certain value. A cutoff of the MTF at 3\% ($\textrm{MTF}_3$) is defined as a limiting resolution that is recognizable with human vision~\cite{langner2009bar,suomalainen2009dosimetry}, and a cutoff of the MTF at 10\% ($\textrm{MTF}_{10}$) is the limiting resolution defined by the International Organization for Standardization (ISO) standard and is used to define the spatial resolution of CT scanners~\cite{smith2003digital,kharfi2012spatial,maier2018medical}. 

For ZnSe(Te), the supercollimating scintillator has a larger area of MTF level between 1\% to 5\% cutoffs, showing visual enhancement at spatial frequencies that can be resolved with the human eye. \Lu~photonic crystal scintillator shows a significant increase in the spatial resolution at $\textrm{MTF}_{10}$, and the enhancement becomes $\sim 10.7 \times$ using the spatial frequency values measured at $\textrm{MTF}_3$. The shape of the PSF and MTF in Fig.~\ref{fig:multiple_frequency}(f) is anisotropic due to the shape of the IFCs. For ZnSe(Te), both the PSF and MTF are relatively isotropic, as supercollimation constitutes only a small part of the overall emission bandwidth. 


Another important metric that captures the performance of the scintillators frequently used in medical imaging is the detector quantum efficiency (DQE). DQE is a spatial frequency-dependent metric defined in terms of the MTF and the noise power spectrum (NPS), where the latter characterizes the spectral distribution of noise in the detected image. Formally, $\textrm{DQE}(f) = \textrm{MTF}^2(f)/(\bar{q}\textrm{NPS}(f))$, where $f$ is the spatial frequency and $\bar{q}$ is the incident X-ray fluence~\cite{beutel2000handbook}.  A system with higher DQE preserves more spatial information while introducing less noise per incident photon, and therefore requires proportionally fewer incident X-ray photons to produce an image of equivalent quality~\cite{don2013image}. The DQE at a high spatial frequency determines the amount of X-ray required to resolve the fine detail of the imaging objects. The right side of Fig.~\ref{fig:multiple_frequency}(c) and (f) clearly demonstrates the significant DQE enhancement of one or two orders of magnitude at high spatial frequencies for both cases.

\section*{X-ray dose reduction in medical imaging}

To demonstrate the potential clinical application of supercollimating photonic crystal scintillators, we simulate X-ray imaging of a human kidney vasculature. X-ray angiography requires high resolution imaging modalities to resolve fine features and to reduce the X-ray dose to the patient. We prepare a kidney sample in our Geant4 simulation using a 3D kidney vasculature reconstructed with Hierarchical Phase-Contrast Tomography~\cite{jain2024vasculature}. We use an iodine contrast agent and X-ray energies near the iodine K-edge to simulate realistic medical imaging conditions with contrast between the vessel and the tissue. Simulated X-ray images are obtained using the \Lu~supercollimating photonic crystal scintillator and a uniform \Lu~scintillator. Details on how we prepare this simulation can be found in the Methods section.

\begin{figure*}
    \centering
    \includegraphics[scale = 1.0]{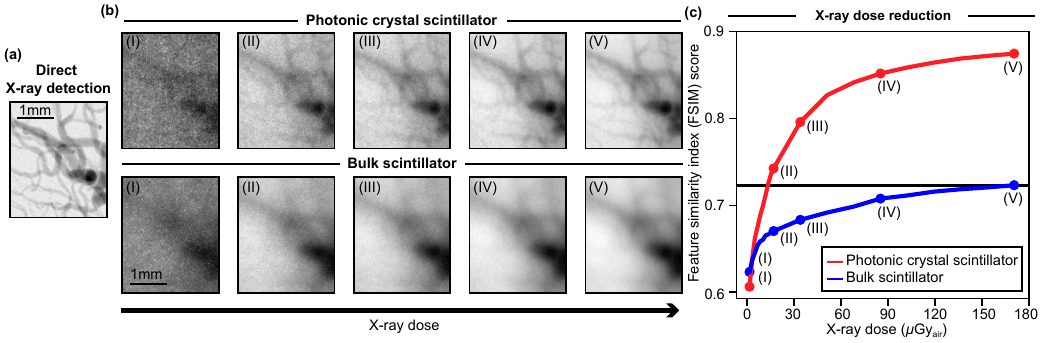}
    \vspace*{-4mm}
    \caption{\textbf{Enhancing medical X-ray imaging performance using supercollimating photonic crystal scintillators.} (\textbf{a}) Direct X-ray detection of a human kidney vasculature (reference). (\textbf{b}) X-ray images of the human kidney vasculature obtained under different X-ray doses with supercollimating photonic crystal scintillators (upper panel) and uniform scintillators (lower panel). The exact X-ray dose of each image is marked in (\textbf{c}). (\textbf{c}) Feature similarity index (FSIM) score dependence as a function of X-ray dose. Supercollimating photonic crystal scintillators can achieve similar FSIM score while using $\sim \times 10$ less X-ray dose. The thickness of the scintillators considered here is \SI{270}{\micro \meter}. For the X-ray image in (\textbf{a}), the incident X-ray dose is set to 10 times higher than the maximum dose used in the scintillator-based images (images in \textbf{(b)}) to compensate for the low detection efficiency of direct X-ray absorption.}
    \label{fig:clinical}
\end{figure*}

Fig.~\ref{fig:clinical}(b) shows X-ray images generated under different X-ray doses. Both the photonic crystal scintillator and the uniform scintillator show poor image quality when the X-ray dose is low, as noise dominates this regime. When the X-ray dose increases, photonic crystal scintillators reveal fine details of the vasculature. Although the uniform scintillator shows an improvement in contrast as the X-ray dose increases, the detailed features of the vessel remain blurred due to lateral spreading of the emitted light. 

This observation points to an important benefit of using supercollimating photonic crystal scintillators---to reduce the X-ray dose without a significant degradation of the image quality. As discussed in Fig.~\ref{fig:multiple_frequency}, imaging systems with a higher DQE can enable this feature. To quantify how the image quality depends on the input X-ray dose, we prepare the reference X-ray image obtained with direct detection (i.e., without scintillators) as shown in the Fig.~\ref{fig:clinical}(a). We then calculate the feature similarity index (FSIM) score of scintillator X-ray images at different X-ray doses~\cite{zhang2011fsim}. Fig.~\ref{fig:clinical}(c) shows that the FSIM score increases faster with supercollimating scintillators. Here we find that to achieve a similar FSIM score to the bulk scintillator, supercollimating photonic crystal scintillators require $\sim 10 \times$ less X-ray dose. In Supplementary Note 3, we simulate X-ray imaging of large vessels of the human kidney using ZnSe(Te) scintillator. The result shows a similar trend as in Fig.~\ref{fig:clinical}, indicating that materials and structures of photonic crystal scintillators can be chosen based on the amount of the spatial resolution enhancement needed when imaging different specimens.

\section{Conclusion and Discussion}
Our work shows that by structuring scintillating materials into three-dimensional photonic crystals with supercollimation, it is possible to substantially improve the spatial resolution for X-ray imaging. This breaks the fundamental tradeoff between the scintillator efficiency and the spatial resolution present in uniform and semi-patterned scintillators. Our results highlight two distinctive advantages by utilizing supercollimating photonic crystal scintillators: (1) enabling high-resolution X-ray imaging beyond conventional schemes, and (2) up to an order of magnitude reduction of X-ray doses while maintaining image quality.

We briefly discuss two approaches to fabricate supercollimating photonic crystal scintillators. One approach is to directly fabricate three-dimensional structures out of scintillator materials, which is possible for certain garnet-based materials using laser nanolithography followed by chemical etching~\cite{rodenas2019three, jurgensen2025volumetrically}. A second approach is to infiltrate scintillator materials into 3D photonic crystal scaffolds made using existing methods such as lithographic assembly~\cite{Subramania2004, qi2004three}, direct laser writing via two-photon polymerization~\cite{deubel2004direct, vaidya2020, schulz2021, chernow2021, peng2016three, chen2019observation, turner2013miniature}, colloidal self-assembly~\cite{wijnhoven1998preparation, lee2024dna, posnjak2024diamond, he2020colloidal}, implosion fabrication~\cite{salamin2026three}, and interference lithography~\cite{moon2006fabricating, divliansky2003fabrication}. To maintain a large refractive index contrast, low-index plastic or liquid scintillators could be infiltrated~\cite{kubo2005control,zhang2019electrically}. Besides the gyroid structures, several other designs are known to support supercollimation, the most general situation arising when the curvature of the IFC flips between two high-symmetry points as a function of frequency, passing through a flat region~\cite{rakich2006achieving, lu2006experimental, iliew2005self, joannopoulos2008molding}.

Beyond the specific structures considered in this work, our framework can be generalized to a broader class of photonic crystal scintillators supporting flat or weakly curved IFCs. The ability to engineer scintillation emission through three-dimensional photonic bandstructure design opens additional opportunities to optimize detector performance, including tailoring angular emission profiles and designing application-specific responses across different X-ray energies and imaging modalities. While the present work focuses on the medical X-ray imaging context, the same principles could be extended to other scintillation-based systems such as high-energy particle detectors. More broadly, our results establish nanophotonic dispersion engineering as a route towards overcoming conventional transport limits in scintillators, enabling detector architectures that simultaneously achieve high efficiency and high spatial resolution.

\section{Materials and methods}
In this section, we briefly describe the numerical framework used to simulate X-ray imaging with supercollimating photonic crystal scintillators. Although Geant4 is a powerful Monte Carlo simulation tool for scintillation and X-ray transport, it does not intrinsically capture the nanophotonic effects introduced by photonic crystal structures. To model these effects, we integrate MIT Photonic Bands (MPB) simulations with Geant4. A more detailed description of the simulation workflow is provided in Supplementary Note 2.

\subsection*{MPB simulation}
For the photonic crystal band structure simulations, we consider two different single gyroid lattice air-hole photonic crystals, implemented in ZnSe(Te) and \Lu~scintillating materials. The filling factor of the gyroid structures are optimized to maximize the frequency bandwidth of two fully-closed IFCs (i.e., gray shaded area shown in Fig.~\ref{fig:single_frequency}(a)). The IFCs of the first two bands are obtained by solving the dispersion relationship. We also sample the group velocity along the IFCs to determine the sampling weight in Equation~\ref{eq:coupling rate}. The angular emission distribution is then sampled and incorporated into the Geant4 simulation framework to model supercollimation in photonic crystal scintillators.   

Another important consideration is the effective scintillation yield of the photonic crystal scintillator. A photonic crystal scintillator with the same thickness as a bulk scintillator contains less scintillating material due to the presence of air. In addition, the density of states (DOS) of the photonic crystal also affects the scintillation performance. Therefore, we calculate the effective scintillation yield that reflects the change in the intrinsic scintillation yield due to the filling factor and the DOS. A detailed discussion can be found in the Supplementary Note 2.

\subsection*{Geant4 simulation}
Geant4 computes the distribution of X-ray energy deposition and the corresponding spatial distribution of dipoles within the scintillator by incorporating key material parameters including X-ray attenuation coefficient, material density, and the scintillator thickness. The number of visible photons generated at each dipole is proportional to the amount of energy deposited. For each photon, the direction of emission is sampled from the angular emission obtained by the nanophotonic simulations from MPB. Subsequently, the emitted light is incoherently summed on the detector plane to acquire the X-ray images. We assume that an ideal silicon detector (i.e., 100\% quantum efficiency) is directly glued on the back-plane of the scintillator. The density of each scintillating material is obtained from Ref.~\cite{sengupta2015bright,linardatos2020optical} and the X-ray attenuation coefficient is computed internally in the Geant4 based on the elemental mass attenuation coefficients~\cite{jackson1981x}. 

For imaging and spatial resolution characterization, an ideal pencil beam with an infinitesimal cross-sectional area is used to measure the PSF. For the MIT logo image shown in the left panel of Fig.~\ref{fig:single_frequency}(b), X-rays pass through the white pixels while they are blocked through the black pixels. The X-ray energy is fixed to \SI{30}{\kilo\eV} for all relevant simulations. For the medical imaging simulations, we create a 3D voxelized human kidney vasculature~\cite{jain2024vasculature} in Geant4. The original vasculature structure was imaged with hierarchical phase-contrast tomography, and the structure is segmented to either blood vessel or tissue. The element composition and the density of two materials are modeled based on the ICRP 110 human phantom data~\cite{menzel2009icrp}. To enhance the contrast between the vessel and the tissue, a standard iodine contrast agent is assumed to be present in the blood vessel with its concentration close to the level used during the ex-vivo imaging. For this situation, we use an X-ray energy of \SI{33.5}{\kilo\eV}, which is positioned near the K-edge of the Iodine.

\section{Data and code availability statement}

All the data and codes that are used within this paper are available from the corresponding authors upon request. Correspondence and requests should be addressed to S.~C. (seouc130@mit.edu) and S.~V. (svaidya1@mit.edu).

\section{Authors contributions}
S.~V., C.~R.-C., and M.~S. conceived the original idea. S.~C. and S.~V. developed a theory for supercollimation in photonic crystal scintillators. S.~C. developed the framework for X-ray imaging with scintillators. S.~C. acquired and analyzed the simulation data with contributions from S.~V., and C.~R.-C.; M.~S. supervised the project. The manuscript was written by S.~C., and S.~V., with inputs from all authors.

\section{Competing interests}

S.~C., S.~V., C.~R.-C., and M.~S. are seeking patent protection for ideas in this work (US provisional patent application no. 63/960,449).

\section{Acknowledgements}

S.~V. dedicates this work to the memory of Ujwala Joshi. We thank S. Min, S. Pajovic, J. Chen, and W. Michaels for stimulating discussion. S.~C. acknowledges support from Korea Foundation for Advanced Studies Overseas PhD Scholarship. C.~R.-C. is supported by a Stanford Science Fellowship. The authors acknowledge the MIT SuperCloud and Lincoln Laboratory Supercomputing Center for providing computation resources supporting the results reported within this paper. This material is based upon work also supported in part by the U.S. Army Research Office through the Institute for Soldier Nanotechnologies at MIT, under Collaborative Agreement Number W911NF-23-2-0121. This work was also supported in part by the DARPA Agreement No. HR00112530318.

\bibliographystyle{unsrt}

\bibliography{bibliography.bib}

@book{gektin2017inorganic,
  title={Inorganic scintillators for detector systems},
  author={Gektin, Alexander and Korzhik, Mikhail},
  year={2017},
  publisher={Springer}
}

@article{moses2002current,
  title={Current trends in scintillator detectors and materials},
  author={Moses, William W},
  journal={Nuclear Instruments and Methods in Physics Research Section A: Accelerators, Spectrometers, Detectors and Associated Equipment},
  volume={487},
  number={1-2},
  pages={123--128},
  year={2002},
  publisher={Elsevier}
}

@article{van2002inorganic,
  title={Inorganic scintillators in medical imaging},
  author={Van Eijk, Carel WE},
  journal={Physics in medicine \& biology},
  volume={47},
  number={8},
  pages={R85},
  year={2002},
  publisher={IOP Publishing}
}

@article{singh2024bright,
  title={Bright innovations: Review of next-generation advances in scintillator engineering},
  author={Singh, Pallavi and Dosovitskiy, Georgy and Bekenstein, Yehonadav},
  journal={ACS nano},
  volume={18},
  number={22},
  pages={14029--14049},
  year={2024},
  publisher={ACS Publications}
}

@article{carr2024toward,
  title={Toward “super-scintillation” with nanomaterials and nanophotonics},
  author={Carr Delgado, Hamish and Moradifar, Parivash and Chinn, Garry and Levin, Craig S and Dionne, Jennifer A},
  journal={Nanophotonics},
  volume={13},
  number={11},
  pages={1953--1962},
  year={2024},
  publisher={De Gruyter}
}

@article{roques2022framework,
  title={A framework for scintillation in nanophotonics},
  author={Roques-Carmes, Charles and Rivera, Nicholas and Ghorashi, Ali and Kooi, Steven E and Yang, Yi and Lin, Zin and Beroz, Justin and Massuda, Aviram and Sloan, Jamison and Romeo, Nicolas and others},
  journal={Science},
  volume={375},
  number={6583},
  pages={eabm9293},
  year={2022},
  publisher={American Association for the Advancement of Science}
}

@article{martin2025large,
  title={Large-scale self-assembled nanophotonic scintillators for X-ray imaging},
  author={Martin-Monier, Louis and Pajovic, Simo and Abebe, Muluneh G and Chen, Joshua and Vaidya, Sachin and Min, Seokhwan and Choi, Seou and Kooi, Steven E and Maes, Bjorn and Hu, Juejun and others},
  journal={Nature Communications},
  volume={16},
  number={1},
  pages={5750},
  year={2025},
  publisher={Nature Publishing Group UK London}
}

@article{kurman2020photonic,
  title={Photonic-crystal scintillators: Molding the flow of light to enhance X-ray and $\gamma$-ray detection},
  author={Kurman, Yaniv and Shultzman, Avner and Segal, Ohad and Pick, Adi and Kaminer, Ido},
  journal={Physical Review Letters},
  volume={125},
  number={4},
  pages={040801},
  year={2020},
  publisher={APS}
}

@article{ye2024nanoplasmonic,
  title={The Nanoplasmonic Purcell Effect in Ultrafast and High-Light-Yield Perovskite Scintillators},
  author={Ye, Wenzheng and Yong, Zhihua and Go, Michael and Kowal, Dominik and Maddalena, Francesco and Tjahjana, Liliana and Wang, Hong and Arramel, Arramel and Dujardin, Christophe and Birowosuto, Muhammad Danang and others},
  journal={Advanced Materials},
  volume={36},
  number={25},
  pages={2309410},
  year={2024},
  publisher={Wiley Online Library}
}

@article{ye2022enhancing,
  title={Enhancing large-area scintillator detection with photonic crystal cavities},
  author={Ye, Wenzheng and Bizarri, Gregory and Birowosuto, Muhammad Danang and Wong, Liang Jie},
  journal={ACS Photonics},
  volume={9},
  number={12},
  pages={3917--3925},
  year={2022},
  publisher={ACS Publications}
}

@article{kurman2024purcell,
  title={Purcell-enhanced X-ray scintillation},
  author={Kurman, Yaniv and Lahav, Neta and Schuetz, Roman and Shultzman, Avner and Roques-Carmes, Charles and Lifshits, Alon and Zaken, Segev and Lenkiewicz, Tom and Strassberg, Rotem and Be’er, Orr and others},
  journal={Science Advances},
  volume={10},
  number={44},
  pages={eadq6325},
  year={2024},
  publisher={American Association for the Advancement of Science}
}

@article{kosaka1999self,
  title={Self-collimating phenomena in photonic crystals},
  author={Kosaka, Hideo and Kawashima, Takayuki and Tomita, Akihisa and Notomi, Masaya and Tamamura, Toshiaki and Sato, Takashi and Kawakami, Shojiro},
  journal={Applied Physics Letters},
  volume={74},
  number={9},
  pages={1212--1214},
  year={1999},
  publisher={American Institute of Physics}
}

@article{prather2004dispersion,
  title={Dispersion-based optical routing in photonic crystals},
  author={Prather, Dennis W and Shi, Shouyuan and Pustai, David M and Chen, Caihua and Venkataraman, Sriram and Sharkawy, Ahmed and Schneider, Garrett J and Murakowski, Janusz},
  journal={Optics letters},
  volume={29},
  number={1},
  pages={50--52},
  year={2004},
  publisher={Optical Society of America}
}

@article{wang2012photonic,
  title={Photonic crystal self-collimation sensor},
  author={Wang, Yufei and Wang, Hailing and Xue, Qikun and Zheng, Wanhua},
  journal={Optics Express},
  volume={20},
  number={11},
  pages={12111--12118},
  year={2012},
  publisher={Optical Society of America}
}

@article{yasar2021spatially,
  title={Spatially resolved x-ray detection with photonic crystal scintillators},
  author={Yasar, Firat and Kilin, M and Dehdashti, S and Yu, Zongfu and Ma, Zhenqiang and Wang, Zhehui},
  journal={Journal of Applied Physics},
  volume={130},
  number={4},
  year={2021},
  publisher={AIP Publishing}
}

@inproceedings{langner2009bar,
  title={Bar and point test patterns generated by dry-etching for measurement of high spatial resolution in micro-{CT}},
  author={Langner, O and Karolczak, M and Rattmann, G and Kalender, WA},
  booktitle={World Congress on Medical Physics and Biomedical Engineering, September 7-12, 2009, Munich, Germany: Vol. 25/2 Diagnostic Imaging},
  pages={428--431},
  year={2009},
  organization={Springer}
}

@article{suomalainen2009dosimetry,
  title={Dosimetry and image quality of four dental cone beam computed tomography scanners compared with multislice computed tomography scanners},
  author={Suomalainen, Anni and Kiljunen, Timo and Kaser, Y and Peltola, Jaakko and Kortesniemi, Mika},
  journal={Dentomaxillofacial Radiology},
  volume={38},
  number={6},
  pages={367--378},
  year={2009},
  publisher={Oxford University Press}
}

@article{klein1968bandgap,
  title={Bandgap dependence and related features of radiation ionization energies in semiconductors},
  author={Klein, Claude A},
  journal={Journal of Applied Physics},
  volume={39},
  number={4},
  pages={2029--2038},
  year={1968},
  publisher={American Institute of Physics}
}

@article{joannopoulos2008molding,
  title={Molding the flow of light},
  author={Joannopoulos, John D and Johnson, Steven G and Winn, Joshua N and Meade, Robert D},
  journal={Princet. Univ. Press. Princeton, NJ [ua]},
  volume={12},
  pages={33},
  year={2008}
}

@article{johnson2001block,
  title={Block-iterative frequency-domain methods for Maxwell’s equations in a planewave basis},
  author={Johnson, Steven G and Joannopoulos, John D},
  journal={Optics express},
  volume={8},
  number={3},
  pages={173--190},
  year={2001},
  publisher={Optical Society of America}
}

@article{agostinelli2003geant4,
  title={Geant4—a simulation toolkit},
  author={Agostinelli, Sea and Allison, John and Amako, K al and Apostolakis, John and Araujo, Henrique and Arce, Pedro and Asai, Makoto and Axen, D and Banerjee, Swagato and Barrand, GJNI and others},
  journal={Nuclear instruments and methods in physics research section A: Accelerators, Spectrometers, Detectors and Associated Equipment},
  volume={506},
  number={3},
  pages={250--303},
  year={2003},
  publisher={Elsevier}
}

@article{sengupta2015bright,
  title={Bright {Lu2O3}: Eu thin-film scintillators for high-resolution radioluminescence microscopy},
  author={Sengupta, Debanti and Miller, Stuart and Marton, Zsolt and Chin, Frederick and Nagarkar, Vivek and Pratx, Guillem},
  journal={Advanced healthcare materials},
  volume={4},
  number={14},
  pages={2064--2070},
  year={2015},
  publisher={Wiley Online Library}
}

@book{smith2003digital,
  title={Digital signal processing: a practical guide for engineers and scientists},
  author={Smith, Steven},
  year={2003},
  publisher={Newnes}
}

@article{kharfi2012spatial,
  title={Spatial resolution limit study of a {CCD} camera and scintillator based neutron imaging system according to {MTF} determination and analysis},
  author={Kharfi, F and Denden, O and Bourenane, A and Bitam, T and Ali, A},
  journal={Applied Radiation and Isotopes},
  volume={70},
  number={1},
  pages={162--166},
  year={2012},
  publisher={Elsevier}
}

@article{linardatos2020optical,
  title={On the optical response of tellurium activated zinc selenide {ZnSe}: Te single crystal},
  author={Linardatos, Dionysios and Konstantinidis, Anastasios and Valais, Ioannis and Ninos, Konstantinos and Kalyvas, Nektarios and Bakas, Athanasios and Kandarakis, Ioannis and Fountos, George and Michail, Christos},
  journal={Crystals},
  volume={10},
  number={11},
  pages={961},
  year={2020},
  publisher={MDPI}
}

@book{beutel2000handbook,
  title={Handbook of medical imaging},
  author={Beutel, Jacob},
  volume={3},
  year={2000},
  publisher={Spie Press}
}

@article{jain2024vasculature,
  title={Vasculature segmentation in {3D} hierarchical phase-contrast tomography images of human kidneys},
  author={Jain, Yashvardhan and Walsh, Claire L and Yagis, Ekin and Aslani, Shahab and Nandanwar, Sonal and Zhou, Yang and Ha, Juhyung and Gustilo, Katherine S and Brunet, Joseph and Rahmani, Shahrokh and others},
  journal={bioRxiv},
  year={2024}
}

@article{zhang2011fsim,
  title={{FSIM}: A feature similarity index for image quality assessment},
  author={Zhang, Lin and Zhang, Lei and Mou, Xuanqin and Zhang, David},
  journal={IEEE transactions on Image Processing},
  volume={20},
  number={8},
  pages={2378--2386},
  year={2011},
  publisher={IEEE}
}

@article{turner2013miniature,
  title={Miniature chiral beamsplitter based on gyroid photonic crystals},
  author={Turner, Mark D and Saba, Matthias and Zhang, Qiming and Cumming, Benjamin P and Schr{\"o}der-Turk, Gerd E and Gu, Min},
  journal={Nature Photonics},
  volume={7},
  number={10},
  pages={801--805},
  year={2013},
  publisher={Nature Publishing Group UK London}
}

@article{peng2016three,
  title={Three-dimensional single gyroid photonic crystals with a mid-infrared bandgap},
  author={Peng, Siying and Zhang, Runyu and Chen, Valerian H and Khabiboulline, Emil T and Braun, Paul and Atwater, Harry A},
  journal={Acs Photonics},
  volume={3},
  number={6},
  pages={1131--1137},
  year={2016},
  publisher={ACS Publications}
}

@article{kubo2005control,
  title={Control of the optical properties of liquid crystal-infiltrated inverse opal structures using photo irradiation and/or an electric field},
  author={Kubo, Shoichi and Gu, Zhong-Ze and Takahashi, Kazuyuki and Fujishima, Akira and Segawa, Hiroshi and Sato, Osamu},
  journal={Chemistry of Materials},
  volume={17},
  number={9},
  pages={2298--2309},
  year={2005},
  publisher={ACS Publications}
}

@article{zhang2019electrically,
  title={Electrically switchable photonic crystals based on liquid-crystal-infiltrated {TiO2}-inverse opals},
  author={Zhang, Ying and Li, Ke and Su, Fengyu and Cai, Zhongyu and Liu, Jianxun and Wu, Xiaowen and He, Huilin and Yin, Zhen and Wang, Lihong and Wang, Bing and others},
  journal={Optics Express},
  volume={27},
  number={11},
  pages={15391--15398},
  year={2019},
  publisher={Optical Society of America}
}

@article{iliew2005self,
  title={Self-collimation of light in three-dimensional photonic crystals},
  author={Iliew, R and Etrich, C and Lederer, F},
  journal={Optics Express},
  volume={13},
  number={18},
  pages={7076--7085},
  year={2005},
  publisher={Optical Society of America}
}

@article{menzel2009icrp,
  title={{ICRP} Publication 110. Realistic reference phantoms: an {ICRP/ICRU} joint effort. A report of adult reference computational phantoms. Ann},
  author={Menzel, Hans-Georg},
  journal={ICRP},
  volume={39},
  pages={1},
  year={2009}
}

@article{jackson1981x,
  title={X-ray attenuation coefficients of elements and mixtures},
  author={Jackson, Daphne F and Hawkes, David J},
  journal={Physics Reports},
  volume={70},
  number={3},
  pages={169--233},
  year={1981},
  publisher={Elsevier}
}

@article{smith2025projected,
  title={Projected lifetime cancer risks from current computed tomography imaging},
  author={Smith-Bindman, Rebecca and Chu, Philip W and Azman Firdaus, Hana and Stewart, Carly and Malekhedayat, Matthew and Alber, Susan and Bolch, Wesley E and Mahendra, Malini and Berrington de Gonz{\'a}lez, Amy and Miglioretti, Diana L},
  journal={JAMA internal medicine},
  volume={185},
  number={6},
  pages={710--719},
  year={2025}
}

@article{yuan2020directional,
  title={Directional control and enhancement of light output of scintillators by using microlens arrays},
  author={Yuan, Di and Liu, Bo and Zhu, Zhichao and Guo, Yaozhen and Cheng, Chuanwei and Chen, Hong and Gu, Mu and Xu, Mengxuan and Chen, Liang and Liu, Jinliang and others},
  journal={ACS applied materials \& interfaces},
  volume={12},
  number={26},
  pages={29473--29480},
  year={2020},
  publisher={ACS Publications}
}

@article{maier2018medical,
  title={Medical imaging systems: An introductory guide},
  author={Maier, Andreas and Steidl, Stefan and Christlein, Vincent and Hornegger, Joachim},
  year={2018},
  publisher={Springer}
}

@article{purcell1946spontaneous,
  title={Spontaneous emission probabilities at radio frequencies},
  author={Purcell, Edward M},
  journal={Physical Review},
  volume={69},
  number={11--12},
  pages={681},
  year={1946},
  publisher={American Physical Society},
  doi={10.1103/PhysRev.69.674.2}
}

@inproceedings{jurgensen2025volumetrically,
  title={Volumetrically-Patterned Nanophotonic Scintillators},
  author={J{\"u}rgensen, Marius and Vaidya, Sachin and Pajovic, Simo and Gales, John P and Chen, Joshua and Katznelson, Shaul and Kooi, Steven E and Richards, Sion and Braddock, Issy and Armstrong, Chris D and others},
  booktitle={2025 Conference on Lasers and Electro-Optics (CLEO)},
  pages={1--2},
  year={2025},
  organization={IEEE}
}

@article{shultzman2023enhanced,
  title={Enhanced imaging using inverse design of nanophotonic scintillators},
  author={Shultzman, Avner and Segal, Ohad and Kurman, Yaniv and Roques-Carmes, Charles and Kaminer, Ido},
  journal={Advanced Optical Materials},
  volume={11},
  number={8},
  pages={2202318},
  year={2023},
  publisher={Wiley Online Library}
}

@article{yang2023photonic,
  title={Photonic flatband resonances for free-electron radiation},
  author={Yang, Yi and Roques-Carmes, Charles and Kooi, Steven E and Tang, Haoning and Beroz, Justin and Mazur, Eric and Kaminer, Ido and Joannopoulos, John D and Solja{\v{c}}i{\'c}, Marin},
  journal={Nature},
  volume={613},
  number={7942},
  pages={42--47},
  year={2023},
  publisher={Nature Publishing Group UK London}
}

@article{don2013image,
  title={Image gently campaign back to basics initiative: ten steps to help manage radiation dose in pediatric digital radiography},
  author={Don, Steven and MacDougall, Robert and Strauss, Keith and Moore, Quentin T and Goske, Marilyn J and Cohen, Mervyn and Herrmann, Tracy and John, Susan D and Noble, Lauren and Morrison, Greg and others},
  journal={American Journal of Roentgenology},
  volume={200},
  number={5},
  pages={W431--W436},
  year={2013},
  publisher={American Roentgen Ray Society}
}

@article{rodenas2019three,
  title={Three-dimensional femtosecond laser nanolithography of crystals},
  author={R{\'o}denas, Air{\'a}n and Gu, Min and Corrielli, Giacomo and Pai{\`e}, Petra and John, Sajeev and Kar, Ajoy K and Osellame, Roberto},
  journal={Nature Photonics},
  volume={13},
  number={2},
  pages={105--109},
  year={2019},
  publisher={Nature Publishing Group UK London}
}

@article{rakich2006achieving,
  title={Achieving centimetre-scale supercollimation in a large-area two-dimensional photonic crystal},
  author={Rakich, Peter T and Dahlem, Marcus S and Tandon, Sheila and Ibanescu, Mihai and Solja{\v{c}}i{\'c}, Marin and Petrich, Gale S and Joannopoulos, John D and Kolodziejski, Leslie A and Ippen, Erich P},
  journal={Nature materials},
  volume={5},
  number={2},
  pages={93--96},
  year={2006},
  publisher={Nature Publishing Group UK London}
}

@article{lu2006experimental,
  title={Experimental demonstration of self-collimation inside a three-dimensional photonic crystal},
  author={Lu, Zhaolin and Shi, Shouyuan and Murakowski, Janusz A and Schneider, Garrett J and Schuetz, Christopher A and Prather, Dennis W},
  journal={Physical review letters},
  volume={96},
  number={17},
  pages={173902},
  year={2006},
  publisher={APS}
}

@article{Subramania2004,
    author = {Subramania, G. and Lin, S. Y.},
    title = {Fabrication of three-dimensional photonic crystal with alignment based on electron beam lithography},
    journal = {Applied Physics Letters},
    volume = {85},
    number = {21},
    pages = {5037-5039},
    year = {2004},
    month = {11},
    issn = {0003-6951},
    doi = {10.1063/1.1825623},
    url = {https://doi.org/10.1063/1.1825623},
}

@article{qi2004three,
  title={A three-dimensional optical photonic crystal with designed point defects},
  author={Qi, Minghao and Lidorikis, Elefterios and Rakich, Peter T and Johnson, Steven G and Joannopoulos, JD and Ippen, Erich P and Smith, Henry I},
  journal={Nature},
  volume={429},
  number={6991},
  pages={538--542},
  year={2004},
  publisher={Nature Publishing Group UK London}
}

@article{deubel2004direct,
  title={Direct laser writing of three-dimensional photonic-crystal templates for telecommunications},
  author={Deubel, Markus and Von Freymann, Georg and Wegener, Martin and Pereira, Suresh and Busch, Kurt and Soukoulis, Costas M},
  journal={Nature materials},
  volume={3},
  number={7},
  pages={444--447},
  year={2004},
  publisher={Nature Publishing Group UK London}
}

@article{vaidya2020,
  title = {Observation of a Charge-2 Photonic Weyl Point in the Infrared},
  author = {Vaidya, Sachin and Noh, Jiho and Cerjan, Alexander and J\"org, Christina and von Freymann, Georg and Rechtsman, Mikael C.},
  journal = {Phys. Rev. Lett.},
  volume = {125},
  issue = {25},
  pages = {253902},
  numpages = {6},
  year = {2020},
  month = {Dec},
  publisher = {American Physical Society},
  doi = {10.1103/PhysRevLett.125.253902},
  url = {https://link.aps.org/doi/10.1103/PhysRevLett.125.253902}
}

@article{schulz2021,
    author = {Schulz, Julian and Vaidya, Sachin and Jörg, Christina},
    title = {Topological photonics in 3D micro-printed systems},
    journal = {APL Photonics},
    volume = {6},
    number = {8},
    pages = {080901},
    year = {2021},
    month = {08},
    issn = {2378-0967},
    doi = {10.1063/5.0058478},
    url = {https://doi.org/10.1063/5.0058478},
}

@article{chernow2021,
author = {Chernow, Victoria F. and Ng, Ryan C. and Peng, Siying and Atwater, Harry A. and Greer, Julia R.},
title = {Dispersion Mapping in 3-Dimensional Core–Shell Photonic Crystal Lattices Capable of Negative Refraction in the Mid-Infrared},
journal = {Nano Letters},
volume = {21},
number = {21},
pages = {9102-9107},
year = {2021},
doi = {10.1021/acs.nanolett.1c02851},
note ={PMID: 34672602},
URL = {https://doi.org/10.1021/acs.nanolett.1c02851},
}

@article{chen2019observation,
  title={Observation of complete photonic bandgap in low refractive index contrast inverse rod-connected diamond structured chalcogenides},
  author={Chen, Lifeng and Morgan, Katrina A and Alzaidy, Ghada A and Huang, Chung-Che and Ho, Ying-Lung Daniel and Taverne, Mike PC and Zheng, Xu and Ren, Zhong and Feng, Zhuo and Zeimpekis, Ioannis and others},
  journal={ACS Photonics},
  volume={6},
  number={5},
  pages={1248--1254},
  year={2019},
  publisher={ACS Publications}
}

@article{wijnhoven1998preparation,
  title={Preparation of photonic crystals made of air spheres in titania},
  author={Wijnhoven, Judith EGJ and Vos, Willem L},
  journal={Science},
  volume={281},
  number={5378},
  pages={802--804},
  year={1998},
  publisher={American Association for the Advancement of Science}
}

@article{lee2024dna,
  title={DNA origami colloidal crystals: opportunities and challenges},
  author={Lee, Jaewon and Kim, Jangwon and Posnjak, Gregor and Ershova, Anastasia and Hayakawa, Daichi and Shih, William M and Rogers, W Benjamin and Ke, Yonggang and Liedl, Tim and Lee, Seungwoo},
  journal={Nano letters},
  volume={25},
  number={1},
  pages={16--27},
  year={2024},
  publisher={ACS Publications}
}

@article{he2020colloidal,
  title={Colloidal diamond},
  author={He, Mingxin and Gales, Johnathon P and Ducrot, {\'E}tienne and Gong, Zhe and Yi, Gi-Ra and Sacanna, Stefano and Pine, David J},
  journal={Nature},
  volume={585},
  number={7826},
  pages={524--529},
  year={2020},
  publisher={Nature Publishing Group UK London}
}

@article{posnjak2024diamond,
  title={Diamond-lattice photonic crystals assembled from DNA origami},
  author={Posnjak, Gregor and Yin, Xin and Butler, Paul and Bienek, Oliver and Dass, Mihir and Lee, Seungwoo and Sharp, Ian D and Liedl, Tim},
  journal={Science},
  volume={384},
  number={6697},
  pages={781--785},
  year={2024},
  publisher={American Association for the Advancement of Science}
}

@article{salamin2026three,
  title={Three-dimensional nanophotonics with spatially modulated optical properties},
  author={Salamin, Yannick and Yang, Gaojie and Mills, Brian and Grossi Fonseca, Andr{\'e} and Roques-Carmes, Charles and Yang, Quansan and Beroz, Justin and Kooi, Steven E and de Miguel Comella, Marc and Mak, Kiran and others},
  journal={Light: Science \& Applications},
  volume={15},
  number={1},
  pages={145},
  year={2026},
  publisher={Nature Publishing Group UK London}
}

@article{divliansky2003fabrication,
  title={Fabrication of three-dimensional polymer photonic crystal structures using single diffraction element interference lithography},
  author={Divliansky, Ivan and Mayer, Theresa S and Holliday, Kito S and Crespi, Vincent H},
  journal={Applied Physics Letters},
  volume={82},
  number={11},
  pages={1667--1669},
  year={2003},
  publisher={American Institute of Physics}
}

@article{moon2006fabricating,
  title={Fabricating three-dimensional polymeric photonic structures by multi-beam interference lithography},
  author={Moon, Jun Hyuk and Ford, Jamie and Yang, Shu},
  journal={Polymers for Advanced Technologies},
  volume={17},
  number={2},
  pages={83--93},
  year={2006},
  publisher={Wiley Online Library}
}

\end{document}